# High Quality Factor High-Temperature Superconducting Microwave Cavity Development for the Dark Matter Axion Search in a Strong Magnetic Field

Danho Ahn[1,2], Dojun Youm[2], Ohjoon Kwon[2], Woohyun Chung[2] and Yannis K. Semertzidis[1,2*]

[1] Center for Axion and Precision Physics Research, Institution of Basic Science, Daejeon, South Korea
[2] Physics Department, Korea Advanced Institute of Science and Technology, Daejeon, South Korea
* Corresponding Author

## Abstract

We demonstrate a superconducting (SC) microwave (mw) cavity that can accelerate the dark matter search by maintaining superconductivity in a high DC magnetic field. We used high temperature superconductor (HTSC) yttrium barium copper oxide (YBCO) with a phase transition temperature of 90K to prevent SC failure by the magnetic field. Since the direct deposition of HTSC film on the metallic mw cavity is very difficult, we used the commercial HTSC tapes which are flexible metallic tapes coated with HTSC thin films. We fabricated resonating cavity ($f_{TM010}$ ~ 6.89 GHz) with a third of the inner wall covered by YBCO tapes and measured the quality factor (Q factor) at 4K temperature, varying the DC magnetic field from 0 to 8 tesla. There was no significant quality (Q) factor drop and the superconductivity was well maintained even in 8 tesla magnetic field. This implies the possibility of good performance of HTSC mw resonant cavity under a strong magnetic field for axion detection.

## 1. Introduction

Superconducting (SC) microwave (mw) cavities are the key component to organize highly sensitive mw systems due to its much lower power loss than normal metals. The SC cavity has various applications, such as particle accelerators [1-5], material characterization at mw frequencies [6-7], and filters for communication systems [8]. Especially, enhancing the quality (Q) factor of the SC mw cavities is one of the main technical issues in the axion search. [9] The scanning rate in the experiment is proportional to the Q factor of the cavity which collects the photons being converted from axions for large conversion signal. [10-11, 27]

In the axion dark matter search, however, the strong magnetic field is the most preferred physical parameter and most of the known experiments to search axions are using bigger than several tesla (T) of magnetic field, where the superconductivity totally disappears. [10-11, 27] For example, the niobium, the most widely using SC material, becomes normal metal much resistive than SC state after 0.3 T though the Q factor of niobium cavity easily reaches over $10^8$. [26] Some conventional superconductors such as niobium titanium (NbTi), magnesium diboride ($MgO_2$) and niobium tin(NbSn), have the higher second critical fields, respectively, at 12 T, 14

T, and 27 T, but the dippining frequencies of those materials are lower than gigahertz (GHz) range, thus the vortex vibration would create additional surface resistance that will degrade the Q factor drastically. [4, 12, 25]

Nonetheless, it is reported that high-temperature superconductors (HTSC) are well surviving even in tesla order of magnetic field. Representatively, the yttrium barium copper oxide (YBCO), has 90K of its second critical temperature, the extreme higher critical field near a hundred teslas for an in-plane direction of the crystal [4], and also has high depinning frequency. (f > 10 GHz) [4, 12, 25] Many studies [13-16] shows its surface resistance is much lower than the copper [24] that promise the best Q factor among normal metals. Unfortunately, acquiring a three-dimensional cavity structure with YBCO or with similar HTSC material is challenging because the superconductivity of the YBCO is dependent on crystal direction. Mw properties of YBCO films without any well-defined texture show degraded mw performance because the crystal mismatch at the grain boundaries adds much resistance on the surface. [8, 17, 23, 24]

To solve this problem, we have to use the film with good textures. One of the most effective ways to exploit the well deposited YBCO film is using the commercial YBCO tape, because the direct deposition of well-textured YBCO film on the curved inner surface is technically impossible. In this work we use the commercial YBCO tape and its fabrication process, structure, properties are well known. [18-19] Using the HTSC surface of the tape, we designed the polygon cavity which consists of 12 pieces.

In this paper, we will discuss the analysis of the polygon cavity and the mw measurement of the polygon cavity design with a third of the inner wall covered by YBCO tapes. We will show the Q factor measurement data with varying the magnetic field. We measured the $TM_{010}$ and $TM_{011}$ (f ~ 6.89 GHz, 7.23 GHz) Q factors of the polygon cavity inner surface at 4.1K, with the radiofrequency (RF) chain and Keysight network analyzer (N5242 PNA-X), varying the magnitude of the applied DC magnetic field from 0 T to 8 T in the BlueFors cryogenic system (LD400).

## 2. Analysis of the Polygon Cavity Design

In the axion search, the mw cavity should follow the several rules because the interaction between axion and electromagnetic wave increase in certain condition. The test cavity, which we use in this work, also should follow the conditions to evaluate the performance of the cavity. From the inverse Primakoff effect, the scanning rate (df/dt) of the system and the form factor (C) of the certain mode are given by, [27] (B: applied magnetic field, V: volume of the cavity, $Q^L$: loaded quality factor, $T_{sys}$: temperature of the system, $e(x)$: normalized electric field of the resonant mode)

$$\frac{df}{dt} \propto \frac{B^4 V^2 C^2 Q^L}{T_{sys}^2}$$

$$C_{lmn} = \frac{\int d^3x \hat{z} \cdot e(x)}{V \int d^3x |e(x)|^2}$$

when the magnetic field is aligned in z-direction. To enhance the sensitivity of the system in the limited time, the system temperature ($T_{sys}$) should be minimized and the volume of the cavity (V), and the form factor of the cavity (C) should be maximized. Since we use the SC electromagnet with the certain bore radius, the cylindrical shape maximize the volume. Due to the direction of the magnetic field, the TM modes, in which the electric field is aligned in z-direction, will maximize the form factor. (Fig. 2)

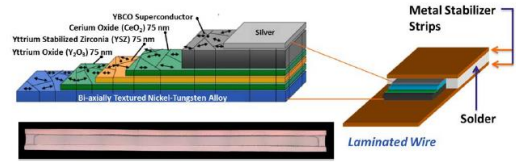

**Figure 1** The architecture of the YBCO tape [18]

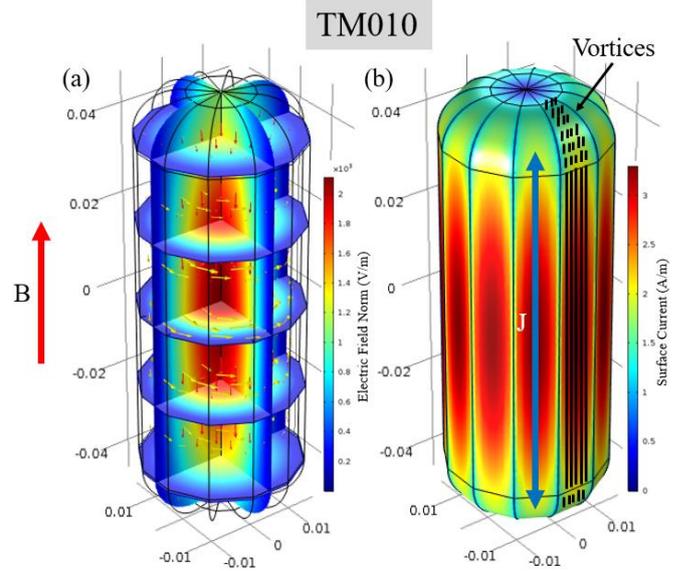

**Figure 2** Simulation result for the $TM_{010}$ mode in the polygon cavity. B is the direction of the applied magnetic field in the axion cavity experiment. (a) The electric field (Red Arrows, Colored 3D Plot) and magnetic field (Yellow Arrows) of the $TM_{010}$ mode, (b) Surface current distribution of the $TM_{010}$ mode. Current flows in the direction of J. The black lines represent vortices in the YBCO film. Since the motion of the vortices is governed by $\mathbf{f} = \mathbf{J} \times \mathbf{\Phi_0}/\mathbf{c}$ ($\mathbf{\Phi_0}$: Magnetic Field Quanta for a Vortex in z-direction), the vortices vibrate at the curved films, but not vibrate at the side wall.

To enhance the sensitivity of the experiment the loaded Q factor ($Q^L$) also should be maximized. SC film is one of the good candidate material for the experiment exhibiting extremely small surface resistance which is inversely proportional to the Q factor. Here we used the YBCO tapes. The film architecture of the YBCO superconducting tape consists of seven parts. [18] On the bi-axially textured nickel-tungsten alloy tape, the 75nm thickness yttrium oxide ($Y_2O_3$) film, the 75nm thickness yttrium stabilized zirconia (YSZ) film, the 75nm thickness cerium oxide ($CeO_2$) film, the 800nm thickness YBCO film and the silver layer for YBCO protection are deposited. Finally, to protect the tape from the mechanical strain, the tape is laminated between two stainless steel tape by soldering. (Fig.1)

However the width of the tapes is limited. To attach the tapes stably, the polygon shape should be introduced. (Fig.2.) The twelve separated pieces are designed for accurate aligning of the YBCO tapes. Since the width of the YBCO tape is 10 mm, the width of the inner surface of each piece is designed as 9 mm. Due to using the TM modes, the vertical cuts of the cylindrical cavity do not make significant degradation of Q factor. However, the sharp edge in the cylinder can be an obstacle to attaching the YBCO tape without any damage on the film. The damage, which is cracks on the film, can give a large additional loss. We replaced angled edge with the rounded edge with radius 10 mm. The modification decrease the form factor of the $TM_{010}$ mode (C ~ 0.5) compared to the form factor of the exact cylindrical cavity (C ~ 0.69). Although, the value is still high compared to the form factor of other modes. The form factor values are obtained from the COMSOL simulations, so we can use this kind of cavity in the axion search.

In point of view of the vortices, there is a qualitative difference between the side walls and the top and bottom rounded surfaces. For TM modes, the direction of the surface current is perpendicular to the magnetic field direction near the surface. The current flow in the line which starts at the top of the cavity inner surface and end at the bottom. Since the motion of the vortices is governed by $\mathbf{f} = \mathbf{J} \times \mathbf{\Phi_0}/c$ ($\mathbf{\Phi_0}$: Magnetic Field Quanta for a Vortex in z-direction), the vortices move at the top and bottom surface, but not move at the side wall. (Fig. 2,3) Since the cross product between the current vector and $\mathbf{\Phi_0}$ is zero at the side wall.

### 3. Cavity Preparation and Measurement

For the measurement the physical quantities of the polygon cavity with YBCO tapes, we expose the HTSC surface of the tapes and attach them on 4 pieces. The four YBCO pieces and the other eight aluminium (Al6061) pieces without YBCO tape attachment are assembled. (Fig. 3)

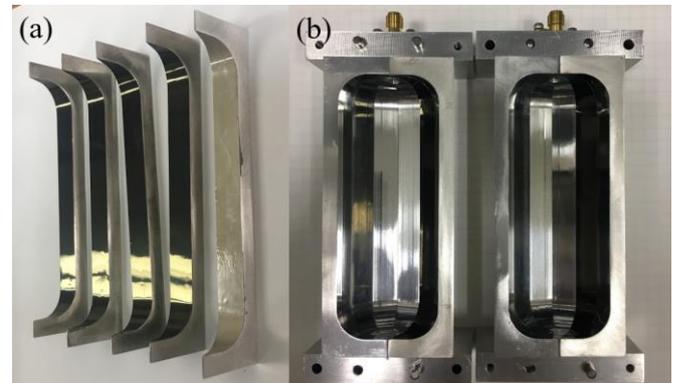

**Figure 3** Cavity Preparation (a) The YBCO film attached cavity pieces, (b) The assembled cavity which consists of the four YBCO tape attached pieces and the eight aluminium pieces.

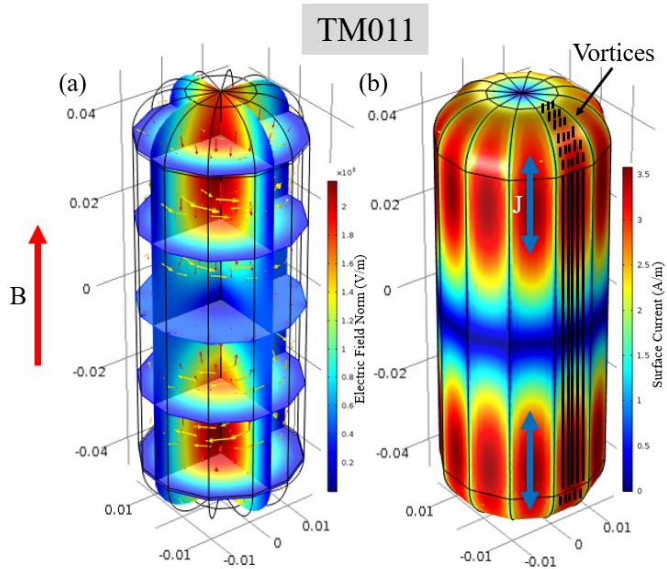

**Figure 3** Simulation result for the $TM_{011}$ mode in the polygon cavity. B is the direction of the applied magnetic field in the axion cavity experiment. (a) The electric field (Red Arrows, Colored 3D Plot) and magnetic field (Yellow Arrows) of the $TM_{010}$ mode, (b) Surface current distribution of the $TM_{010}$ mode. Current flows in the direction of J. The black lines represent vortices in the YBCO film. Since the motion of the vortices is governed by $\mathbf{f} = \mathbf{J} \times \mathbf{\Phi_0}/c$, the vortices vibrate at the curved films, but not vibrate at the side wall.

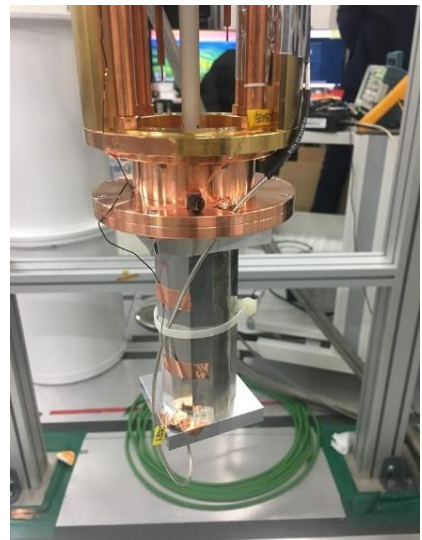

**Figure 4** The polygon cavity installation in the cryocooler. The two RF lines connect to the cavity and two temperature sensor attached on the top and bottom parts of the cavity.

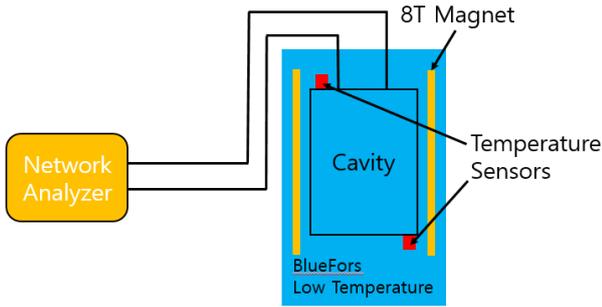

**Figure 5** The schematic of RF chain for the measurement.

After assmbling the parts of the cavity, the cavity installed in the cryogenic system. With the two pole antennas in the cavity connects to the network analyzer to measure the loaded Q factor ($Q^L$), the coupling parameters ($\beta_i$) for two antennas from the measured scattering parameters ($S_{ij}$), and the resonance frequency (f) (Fig. 4,5) The unloaded Q factor ($Q^U$) and the antenna couplings ($\beta_i$) are can be calculated from S parameters ($S_{ij}$). [21]

$$Q^U = (1 + \beta_1 + \beta_2)Q^L$$

$$\beta_i = \left|\frac{1 - S_{ii}}{1 + S_{ii}}\right|$$

Also, we can understand the geometric change of the cavity with the resonance frequency shift. Since the resonance frequency of $TM_{010}$ mode is related to the radius of the cavity, and the frequency of the $TM_{011}$ mode has a relationship with the radius and the height of the cavity. The radius and the height can be increased by thermal shrinkage and decrease of the penetration depth of the YBCO film.

In the cryocooler, the AMI 8 T electromagnet is installed. The center of the magnetic field is at the 400 mm below the bottom of the mixing plate, so the support structure had been designed for aligning the center of the cavity and the magnet center. During measurement of the $TM_{010}$ Q factor, the system temperature had been varied from 300K to 4.1 K. The Q factor measurement of the polygon cavity which consists of twelve aluminium pieces had been conducted at 4.2 K. In the Q factor measurement in the DC magnetic field, the magnetic field was varied from 0 to 8 T.

## 4. Result

From the Q factor measurement for the polygon cavity with the four YBCO pieces with varying temperature, the critical temperature of the YBCO film was characterized at 88K. The surface loss of the YBCO film is calculated as almost zero, because the Q factor of the polygon cavity with four YBCO pieces at 4.1 K was same as the calculated Q factor from the Q factor of the full aluminium cavity at 4.2 K. (Fig. 5) If the surface loss of the four YBCO pieces of cavity fully vanished when it becomes superconductor, the Q factor of the polygon cavity with the four YBCO pieces at 4 K is same as three half of the Q factor of the aluminium cavity. The anomalous frequency change is also observed at the critical temperature. It shows there was the penetration depth change at the temperature. Since thermal shrinkage has no critical behavior at the temperature.

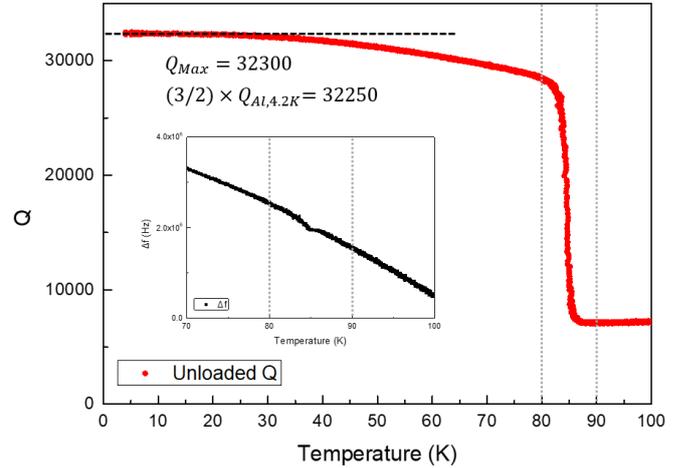

**Figure 5** The Q factor measurement (resonance frequency difference) data of the polygon cavity with the four YBCO pieces from 100 K to 4.1 K (from 100 K to 70 K). The critical temperature of the YBCO film was characterized at 88K. At the temperature, the anomalous change of resonance frequency appeared. The global change of resonance frequency is due to the thermal shrinkage, and the local anomalous change of the frequency is due to the penetration depth change of the superconductor. The Q factor at the 4.1 K is almost the same as two third of the Q factor of the full aluminium cavity. It means YBCO films give much lower surface loss than the aluminium surface.

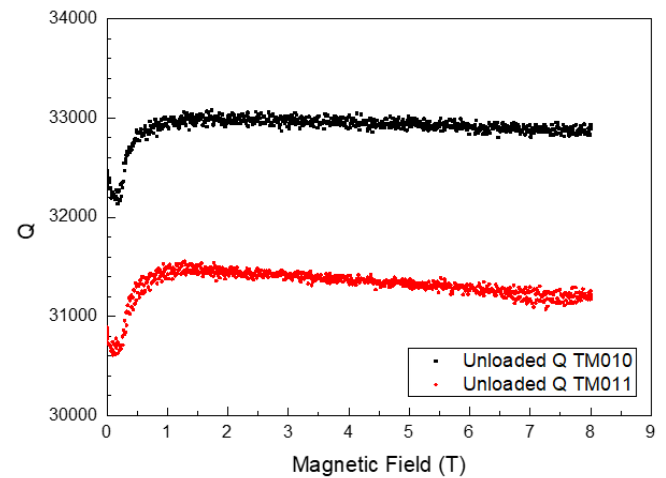

**Figure 6** The Q factor measurement data of the polygon cavity with the four YBCO pieces from 0 to 8 T and vice versa. After increasing Q factor until 1T, the Q factor was decreased gradually. The decreasing slope is larger for the $TM_{011}$ mode which has larger ratio of the top and bottom surface loss.

The Q factor measurement data of the polygon cavity with the four YBCO pieces with varying the magnetic field from 0 T to 8 T shows that the surface loss of the YBCO film was not decreased significantly. The Q factor of the TM010 mode decreased only a few hundred. ($\Delta Q/Q \sim 0.003$) (Fig. 6) Also we can support the data after 1 T is due to the vortex loss from the different decreasing rate of the Q factor between $TM_{010}$ and $TM_{011}$ modes from 1 to 8 T. The result can be explained by the vortex vibration. The Q factor of $TM_{011}$ mode should be decreased faster than the $TM_{010}$ mode, because the surface loss ratios of the top and bottom surfaces are 0.302 for the $TM_{011}$ mode and 0.117 for the $TM_{010}$ mode. The ratios are calculated upon the assumption that all the surface is uniform normal metal by the simulation. However, the vortices at the top and bottom curved surface are vibrated by the resonant modes, but the vortices at the side wall are not vibrated by the resonant modes. (Fig. 2,3) The real difference between the two ratios will bigger than the simulation. In conclusion, we can expect that the Q factor of the polygon cavity with full YBCO pieces will show the high Q factor without any large degradation until 8 T.

## 5. Discussion

In summary, we have designed the polygon cavity for testing $TM_{010}$ modes with the YBCO film to develop the cavity for the dark matter axion search. To implement the YBCO surface on the cavity inner wall, we attached the tapes on the four cavity pieces. After assemblage, we measured the loaded Q factor, the antenna couplings, and the resonance frequency of the cavity for $TM_{010}$ and $TM_{011}$ modes in the cryocooler, using the RF chain and network analyzer. From the temperature variation, we could characterize the critical temperature of YBCO film at 88K, and notice that the surface resistance of the YBCO film is almost zero in the experimental sensitivity. In the Q factor data with varying magnetic field shows that the Q factor value will not decrease significantly until 8 T. The difference of Q factor degradation ratio of two modes support the theory that decreasing Q factor after 1 T is due to the relationship between the vortex direction and surface current direction. Therefore we can expect that the polygon cavity with the full YBCO pieces will show high Q factor in a strong magnetic field.

## References

[1] Padamsee H 2001 Supercond. Sci. Technol. **14** R28
[2] Padamsee H 2017 Supercond. Sci. Technol. **30** 053003
[3] Calatroni S, Bellingeri E, Ferdeghini C, Putti M, Vaglio R, Baumgartner T and Eisterer M 2017 Supercond. Sci. Technol. **30** 075002
[4] Calatroni S 2015 Proc. SRF (Whistler 2015)
[5] Collings E W, Sumption M D and Tajima T 2004 Supercond. Sci. Technol. **17** S595
[6] Zhai Z, Kusko C, Hakim N and Sridhar S 2000 Rev. Sci. Instrum. **71** 3151
[7] Wosik J, Krupka J, Qin K, Ketharnath D, Galstyan E and Selvamanickam V 2017 Spercond. Sci. Technol. **30** 035009
[8] Pandit H, Shi D, Babu N H, Chaud X, Cardwell D A, He P, Isfort D, Tournier R, Mast D and Ferendeci A M 2005 Physica C **425** 44
[9] Sikivie P 2013 arXiv:1009.0762v2
[10] Sikivie P 1983 Phys. Rev Lett. **51**, 1415
[11] Sikivie P 1985 Phys. Rev. D **32**, 2988
[12] Golosovsky M, Tsindlekht M, Chayet H and Davidov D 1994 Phy. Rev. B **50** 470
[13] Sato K, Sato S, Ichikawa K, Watanabe M, Honma T, Tanaka Y, Oikawa S, Saito A and Ohshima S 2014 J. Phys.: Conf. Ser. **507** 012045
[14] Honna T, Sato S, Sato K, Watanabe M, Saito A, Koike K, Kato H and Ohshima S 2013 Physica C **484** 46
[15] Ohshima S, Kitamura K, Noguchi Y, Sekiya N, Saito A, Hirano S and Okai D 2006 J. Phys.: Conf. Ser. **43** 551
[16] Ohshima S, Shirakawa M, Kitamura K, Saito A, Ihara H and Tanaka Y 2004 Chi. J. Phys. **42** 425
[17] Golosovsky M 1998 Part. Accel. **351** 87
[18] Li X, Rupich M W, Thieme C L H, Teplitsky M, Sathyamurthy S, Thompson E, Siegal E, Buczek D, Schreiber J, DeMoranville K, Hannus D, Lynch J, Inch J, Tucker D, Savoy R and Fleshler S 2009 IEEE Trans. Appl. Supercond. **19** 3231
[19] Rupich M W, Li X, Sathyamurthy S, Thieme C L H, DeMoranville K, Gannon J and Fleshler S 2013 IEEE Trans. Appl. Supercond. **23** 6601205
[20] Youm D 2018 Annual Report CAPP
[21] Lancaster M J 1997 *Passive Microwave Device Applications of High-Temperature Superconductors* (Cambridge: Cambridge University Press)
[22] Cahill A, Fukasawa A, Rosenzweig J, Bowden G B, Dolgashev V A, Guo J, Franzi M, Tantawi S, Welander P B and Yoneda C 2016 Proc. IPAC (Busan 2016)
[23] Hein M 1999 *High-Temperature Superconductor Thin Films at Microwave Frequencies (Springer Tracts of Modern Physics vol 155)* (Heidelberg: Springer)
[24] Zahopoulos C, Kennedy W L and Sridhar S 1988 Appl. Phys. Lett. **52** 2168
[25] Golosovsky M, Tsindlekht M and Davidov D 1996 Supercond. Sci. Technol. **9** 1-15
[26] Karasik V R and Shebalin I Y 1970 Sov. Phys. JETP **30**, 1068
[27] Brubaker M B 2016 Dissertation Yale University